\documentstyle{article}
\begin{document}

\begin{titlepage}
\flushright{\bf IPGAS-HEP-TH/43/95\\
            December, 1995     }

\vspace{3cm}

\begin{center}
{\bf NATURAL SUPPRESSION OF~ d~=~5~ OPERATOR INDUCED PROTON DECAY \\
IN SUPERSYMMETRIC  GRAND UNIFIED THEORIES}

\vspace{1cm}

I.~Gogoladze~\footnote {E-mail:
ilia@physics.iberiapac.ge} and~ A.~Kobakhidze~\footnote{E-mail:
yoko@physics.iberiapac.ge}

{\sl Institute of Physics, Georgian Academy of Sciences  }

{\sl Tamarashvili 6, 380077 Tbilisi, Georgia }
\end{center}

\begin{abstract}
The proton decay in supersymmetric (SUSY) Grand Unified Theories (GUTs),
where the special mechanism provides the natural suppression of the
dimension 5 ($d=5$) operators, was studied. Due to this mechanism, $d=5$
operator induced proton decay takes place only through the mixing of light
generation fermions with those from third generation and crucially depends
on the structure of fermion mass matrices. We show that in general proton
decay is strongly suppressed relative to the standard SUSY GUT situation
and for wide class of fermion mass matrices the SUSY parameter space
practically has no restrictions.
\end{abstract}

\vspace{7cm}

\centerline{\Large \bf To be published in J.of Atomic Nuclei
(Yad.Fiz)}
\end{titlepage}

\section{\bf Introduction}

In the Grand Unified Theories (GUTs) supersymmetry introduces a qualitatively
new mechanism of proton decay through the $d=5$ operators \cite{1}. This
baryon number violating operators are induced by color triplet Higgsino
exchange and have the form of F-product
 \begin{equation}
  \frac{1}{M_{H_c}} \biggl( QQQL \biggr)_F
\end{equation}
where $Q,\,L$ denote quark and lepton superfields respectively and $M_{H_c}$
is Higgsino mass. $d=5$ operators convert the fermionic (quark and lepton)
states into the scalar ones (squarks, sleptons). So to be relevant for
the proton decay, the later must be "once again" converted back to the fermionic
states via gaugino (chargino, gluino, ...) exchange.  However, since the
masses of $SU(3)_c \otimes SU(2)_L \otimes U(1)$ gauginos are attributed to
the supersymmetry treating scale ($\sim$ 100 GeV), the resulting four
fermionic operator is suppressed by single power of inverse Higgsino mass
$M^{-1}_{H_c}$ (which at best $\sim M_{CUT}^{-1}$).

For this reason $d=5$ operators give an unacceptable short proton life-time,
unless $M_{H_c} \geq 10^{17}$ GeV \cite{2} and the supersymmetric
spectrum is not too light.  In fact, demanding $M_{H_c} \leq
M_{GUT}$, so that the Yukawa coupling generating the $\tilde{H_c}$ mass
remains perturbative and naturalness criterion,
$m_{\tilde {q},\,\tilde {l}} \leq$ 1 TeV it has been shown \cite{3} that
the proton decay mode $p\rightarrow \bar \nu K^+$ is sufficiently
suppressed only if the squarks and sleptons are heavy, the two lightest
neutralinos and the lightest chargino are much lighter, and $tan\beta\leq
6$. Thus, acceptable Higgsino induced $d=5$ operators are allowed only in
some region of the theory parameters and their suppression is an
important problem in SUSY GUTs.

The natural solution of the problem was proposed in \cite{4}, where it
was shown that colored triplet Higgs (Higgsino) can be automatically
decoupled from the light matter superfield if the quark and lepton masses
are induced trough the higher dimentional operators involving GUT
symmetry violating VEVs.  The mechanism of \cite{4} can be easily
demonstrated on the $SO(10)$ example in which electroweak Higgs doublet
(and color triplet) are extracted from the 10-dimentional representation
and quarks and leptons are placed in spinorial 16-plets. Imagine, that
the masses of light matter generations are induced trough the "Yukawa"
couplings of the form \begin{equation} \frac{1}{M_{GUT}}10_i 45_{ik}
16\gamma_k 16 \end{equation} (only $SO(10)$ tensor indices $i,k$=1,...,10
are written explicitly). Here 45-plet is one of the Higgs representations
that breaks GUT symmetry.  Coupling (2) should be treated as an effective
operator generated trough the heavy (with mass $\sim M_{GUT}$)
$144+\overline {144}$-plet exchange (for more details see [4]).

Obviously, the colored Higgs (Higgsino) automatically gets decoupled from
matter superfields, provided 45-plet develops VEV of the following form
\begin{equation}
       <45>=diag(0,0,0,\sigma,\sigma)V, \hspace{0.5 cm} V \sim M_{GUT},
\end{equation}
where
\begin{displaymath}
{\sigma}=
\left( \begin{array}{cc}
 0 & 1 \\
-1 & 0
\end{array} \right)
\end{displaymath}

Consequently there are no tree level $d=5$ as well as $d=6$
baryon number violating operators.

In the present paper we generalize this mechanism for the "realistic"
fermion mass matrix framework. That is we make natural assumption that only
the third family fermions (or perhaps only the top quark) get masses from the tree level
coupling with 10-plet
\begin{equation}
10 \cdot 16^3 \cdot 16^3
\end{equation}
Model dependently the theory can be arranged in such a way that all third
family (top, bottom and $\tau$) or only top acquire mass from above
operator depending whether both "up" and "down" electroweak Higss
doublets ($H_u$, $H_d$) or only $H_u$ reside in 10-plet. Alternatively,
some intermediate cases are also possible in which $H_d$ only
partially resides in 10-plet and partially in some other
representation that is not coupled with $16^3$.

In general case other fermions will get their masses from the set of
higher dimensional operators of the form
\begin{equation}
O^{\alpha \beta} = \frac{1}{M^{n+m}} (16^{\alpha} 45^m 16^{\beta})_k 10_i
(45^n)_{ik}
\end{equation}
where $(45^n)_{ik}$ denotes an $n$-linear product of
Higgs 45-plets with all other indices (except i, k) contracted among each
other and $\alpha, \beta$ are generation indeces. The regulator scale M
that stands in denominator need not be treated necessarily as an
$SO(10)$-invariant mass, since the contribution to the masses of the
"integrated out" heavy states (e.g. $144+\overline {144}, 16+\overline
{16}$) comes from the large GUT symmetry breaking VEVs as well. The
detailed structure of the operators (4) is not important for
us~\footnote {The most general analysis of such operators in view of
maximal predictivity was done in \cite{5}}. We simply assume that this
structure provides a realistic spectrum of the matter fermions.
Following \cite{4}, the key assumption
is that the operators (5) that give the masses to light fermions enter the
10-plet in the combination
\begin{equation}
10_i (45^n)_{1k} ,
\end{equation}
where at least one 45 has the VEV given by (3).
Thus in our case the colored
triplet Higgs (Higgsino) is coupled only to the "initial" third family
$16^3$.  Since in general this particles are not mass eigenstates, the
coupling of colored Higgs (Higgsino) with physical u,d-quark (squark) can
arise through the mixing in the fermion mass matrices and consequently
will be suppressed by corresponding mixing angles. So proton decay amplitude
will crucially depend on the structure of the fermion mass matrices.

\section{\bf Nucleon decay for various mass matrices.}

The four-fermion baryon number violating chargino dressing effective
Lagrangian can be written down as \cite{1} (gluino and neutralino dressed
contributions are
negligible in comparison with chargino ones)
\begin{eqnarray}
L = \frac{\alpha_2}{2\pi M_{H_c}}\, g_{ii}^u\, g_{kk}^d\,
V_{jk}^{*}\, A_S\, A_L\times
\qquad\qquad\qquad\qquad\qquad\qquad\qquad\qquad\qquad \nonumber    \\
 \times \biggl( (u'_i d'_i)(d'_j\nu'_k)[F(\tilde{u}_j,\tilde{e}_k)+
F(\tilde{u}_i,\tilde{d_i})]
+(d'_i,u'_i)(u'_i,e'_k)[F(\tilde{u_i},\tilde{d_i})+
F(\tilde{d_j},\tilde{\nu_k})]+ \nonumber \\
+(d'_i,\nu'_k)(d'_i,u'_j)[F(\tilde{u}_i,\tilde{e}_k)
+F(\tilde{u}_i,\tilde{d_j})]+
(u'_i d'_j)(u'_i e'_k)
[F(\tilde{d_j},\tilde{u}_i)+F(\tilde{d_j},\tilde{\nu_k})]\biggr),
\end{eqnarray}
where $u'_i,d'_i,e'_i,\nu'_i$ are up and down quarks and leptons mass
eigenstates;
$$F(\tilde{a},\tilde{b})=
\frac{m_{\tilde W}}{m_{\tilde a}^2-m_{\tilde b}^2}
\biggl(\frac{m_{\tilde a}^2}{m_{\tilde a}^2-m_{\tilde W}^2}\,
ln \frac{m_{\tilde a}^2}{m_{\tilde W}^2}-
\frac{m_{\tilde b}^2}{m_{\tilde b}^2-m_{\tilde W}^2}\,
ln \frac{m_{\tilde b}^2}{m_{\tilde W}^2}) \biggr)$$
comes from the loop integral, and $m_{\tilde{a},\,\tilde{b}}$ are s-fermion
(squark, slepton) masses ; $A_L\approx 0.283,\, A_S\approx 0.883$ are the
long-range, short-range RGE suppression factors; $g_{ij}^u$ and $g_{ij}^d $
are Yukawa coupling constants for up and down quarks respectively; $V_{ij}$
are Cabibbo-Kobaiashi-Maskawa (CKM) matrix elements.

\vspace{0.5 cm}

a. {\it The diagonal up-quark mass matrix case }.

Let us consider the case when up-quark mass  matrix has the diagonal
form. In this case quark mixing (in the weak current) entirely arises
 from the down-quark mass matrix $M_d$ which is diagonalized by
CKM rotation. Since Higgs colored triplet is
initially coupled only with the third generation, one of the squarks
participating in $d=5$ operator should be s-top. Consequently, dominant
nucleon decay modes are $p\rightarrow K^+{\bar{\nu}_{\tau}}$, $n\rightarrow
K^0{\bar{\nu}_{\tau}}$ through the  diagram of Fig.1
\begin{figure}[t]
\begin{center}
\setlength{\unitlength}{0.5mm}
\begin{picture}(40,40)
\put(0,0){\line(1,0){20}}
\put(20,0){\line(1,0){20}}
\put(41,0){\line(1,0){3}}
\put(47,0){\line(1,0){3}}
\put(53,0){\line(1,0){3}}
\put(59,0){\line(1,0){3}}
\put(65,0){\line(1,0){3}}
\put(71,0){\line(1,0){3}}
\put(77,0){\line(1,0){3}}
\put(83,0){\line(1,0){3}}
\put(121,0){\line(-1,0){20}}
\put(81,0){\line(1,0){20}}
\put(0,40){\line(1,0){20}}
\put(20,40){\line(1,0){20}}
\put(41,40){\line(1,0){3}}
\put(47,40){\line(1,0){3}}
\put(53,40){\line(1,0){3}}
\put(59,40){\line(1,0){3}}
\put(65,40){\line(1,0){3}}
\put(71,40){\line(1,0){3}}
\put(77,40){\line(1,0){3}}
\put(83,40){\line(1,0){3}}
\put(121,40){\line(-1,0){20}}
\put(81,40){\line(1,0){20}}
\put(40,0){\line(0,1){40}}
\put(81,0){\line(0,1){40}}
\put(10,45){\makebox(0,0){$u$$_L$}}
\put(115,45){\makebox(0,0){$\nu_\tau $$_L$}}
\put(115,-5){\makebox(0,0){$s$$_L$}}
\put(10,-5){\makebox(0,0){$d$$_L$}}
\put(40,20){\makebox(0,0){$\times$}}
\put(81,20){\makebox(0,0){$\times$}}
\put(75,20){\makebox(0,0){$\tilde H_c$}}
\put(35,20){\makebox(0,0){$\tilde W$}}
\put(60,45){\makebox(0,0){$\tilde b$$_L$}}
\put(60,-5){\makebox(0,0){$\tilde t$$_L$}}
\put(20,0){\makebox(0,0){$>$}}
\put(20,40){\makebox(0,0){$>$}}
\put(101,0){\makebox(0,0){$<$}}
\put(101,40){\makebox(0,0){$<$}}
\put(0,-15){\line(1,0){121}}
\put(60,-15){\makebox(0,0){$>$}}
\put(60,-23){\makebox(0,0){$u$$_L$$\,\,$$($$
d$$_L$$)$}}
\end{picture}
\end{center}

\vspace{0.7cm}

\caption{The nucleon decay diagram in the case of diagonal up-quark
mass matrix.}
\end{figure}
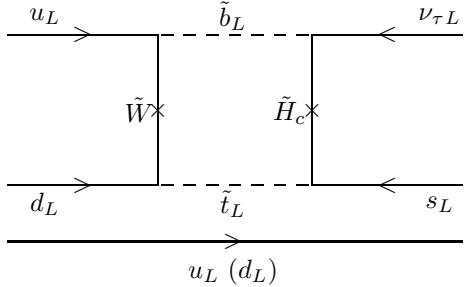

To calculate the life time of proton and neutron, one employs the chiral  Lagrangian
approach to parametrize the hadronic matrix elements \cite{6}. We have
\begin{equation}
 \Gamma (p \rightarrow K^+{\bar{\nu}_{\tau}}) \approx \Gamma (n\rightarrow
K^0{\bar{\nu}_{\tau}})=\frac{\beta^2}{M_{H_C}^2}\frac{m_N}{32 \pi f^2_{\pi}}
\biggl[ 1-\frac{m_K^2}{m_N^2} \biggr]^2(A_L A_S)^2 \mid C_{\nu_{\tau}K}
\mid ^2,
\end{equation}
where $f_{\pi}$=139 MeV is the pion decay constant, and $m_N$=0.938 MeV
and $m_K$=0.439 GeV are the nucleon and K-meson masses, respectively.
The parameter
$\beta$ is the three-quark matrix element of the nucleon wave function
$U_u^{\gamma}$ ($\gamma$=1,2): $\beta U_L^{\gamma}=\epsilon_{abc} <0 \mid
\epsilon_{\alpha \beta} d^{\alpha}_{aL} U^{\beta}_{bL} U^{\gamma}_{cL}\mid
p>$.  Lattice gauge theory gives $\beta=(5.6\pm 0.8)*10^{-3}$ $GeV^3$
\cite{7}.  The parameter $C_{\nu_{\tau}K}$ has the form \begin{equation}
C_{\nu_{\tau K}}=\frac {\alpha_2^2}{2M_W^2 \sin{2\beta}} m_b m_t V_{31}^+
V_{13} V_{23} (V_{23} V_{33}) F(\tilde t,\tilde b) \end{equation} Note that
in a sharp contrast with the standard SUSY GUTs of ref. [1] (in which
Yukawa couplings of the weak doublet and color triplet are not splitted) we
have an additional suppression factor $(V_{23}V_{33})$ in
$C_{\nu_{\tau}K}$.

Using the above values, we have obtained the partial life-time
for $p\rightarrow K^+{\bar{\nu}_{\tau}}$ decay:
\begin{equation}
 \tau(p\rightarrow K^+{\bar{\nu}_{\tau}}) \approx \tau(n\rightarrow
K^0{\bar{\nu}_{\tau}})\approx 7\cdot 10^{-12}\frac
{M_{H_C}^2}{\beta^2}\frac {1}{(V_{23}V_{33})^2}\biggl(\frac
{\sin{2\beta}}{F(\tilde t,\tilde b)}\biggr)^2\quad yr
\end{equation}

In Fig.2 and Fig.3 we have plotted the value of proton partial life-time
$\tau_p (p\rightarrow {\bar{\nu}_{\tau}}K^+)$ as a function of $tan\beta$
and correlations between chargino $m_{\tilde g}$ and squarks $m_{\tilde q}$
masses respectively for various F. One can see from Fig.2 that, in
contrast with usual SUSY GUTs \cite{1}, \cite{2}, the wide range of
$tan\beta$ parameter is allowed. From Fig.3 it is clear, that the large
splitting between chargino and squark masses is not necessary, so in our
case the allowed SUSY parameter space is much larger than in usual SUSY
GUTs.
\begin{figure}[b]
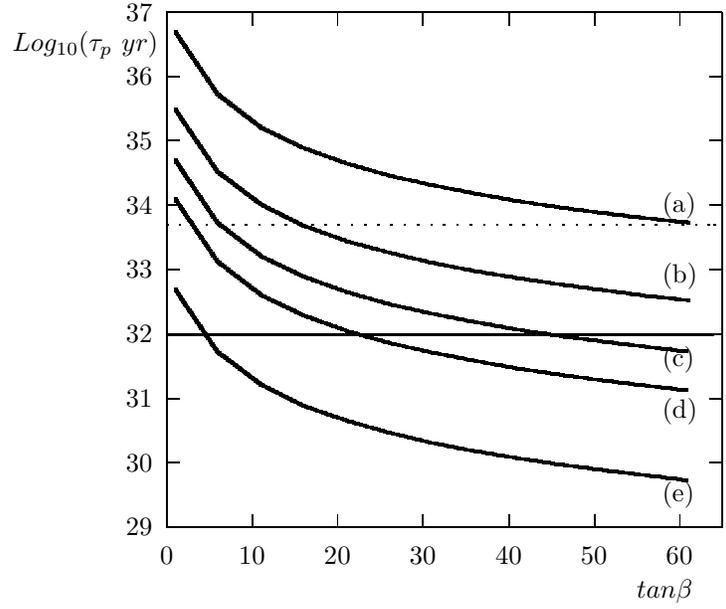

\begin{center}
\setlength{\unitlength}{0.240900pt}
\ifx\plotpoint\undefined\newsavebox{\plotpoint}\fi
\sbox{\plotpoint}{\rule[-0.175pt]{0.350pt}{0.350pt}}%

\end{center}

\vspace{-0.5cm}

\caption{{\small Value of proton partial life-time
$\tau_p(p\rightarrow \bar {\nu}_\tau K^+)$ as function of $tan\beta$
for various F. The (a), (b), (c), (d), (e) lines correspond to $F=
5\cdot 10^{-5},\,\,2\cdot 10^{-4},\,\,5\cdot
10^{-4},\,\,10^{-3}\,\,and\,\,0.05$.  The solid horizontal
line is the current experimental limit. The dotted line is the upper
bound for Super-Kamiokande.}}
\end{figure}
\begin{figure}[b]
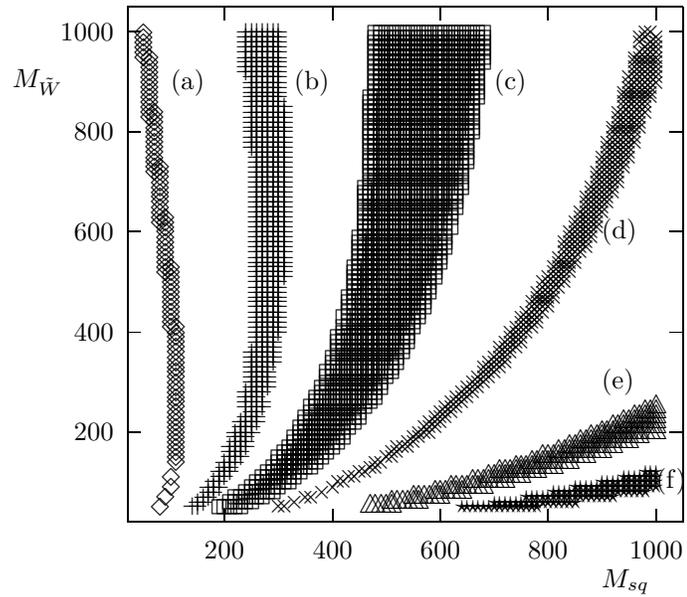

\begin{center}
\setlength{\unitlength}{0.240900pt}
\ifx\plotpoint\undefined\newsavebox{\plotpoint}\fi
\sbox{\plotpoint}{\rule[-0.175pt]{0.350pt}{0.350pt}}%

\end{center}

\vspace{0.5cm}

\caption{Correlations between chargino and squark masses. The  (a), (b), (c),
(d), (e), (f) points correspond to $F= 0.05,\,\, 0.02,\,\, 10^{-3},\,\,
5\cdot 10^{-4},\,\, 2\cdot 10^{-4}\, and\, 10^{-4}$.}
\end{figure}

\vspace{0.5 cm}

b. {\it The diagonal down-quark mass matrix case.}

Now, let us consider the case of the diagonal down-quark and charged
lepton mass matrix.  Then in the limit in which quark and squark masses are
diagonalized simultaneously, there are no gagino dressed diagrams for proton
decay. The incoming pair of the constituent quarks, participating in the
diagram must be u-u or d-d or  u-d. In the later case the one of the
outgoing legs from the diagram (due to charge and color conservation) should
be down type quark or charged lepton.  But the only down quark (charged
lepton) coupled to triplet Higgsino is b-quark ($\tau$-lepton).  Thus,
proton decay is kinematically forbidden.

If the incoming legs in the diagram are u-u, those the outgoing  should be
down-quark + charged lepton. Consequently, this diagram is dressed either by
gaugino or neutralino, which do not change flavor. Once again we inevitably
end up with b-quark bieng one of the outgoing legs of the diagram and thus
proton decay is impossible. The same arguments can be used to show that
there are no $d=5$ diagrams responsible for neutron decay in
the case when incoming legs are d-d.

It is clear, that in any
mass matrix ansatzes, where the first generation down-type quark does not mix
with the third one, the situation is similar to one discussed above

In conclusion, it should be stressed that we have considered the examples
a) and b) in which one of the mass matrices are diagonal just for
simplisity and maximal predictivity, since in these cases suppression
factors depend on CKM matrix elements only. In the case of more general
anzats (with both matrices non-diagonal) the analogous suppression of
proton decay will take place. The only difference is that in this case
suppression factor will depend on the less known quark mass matrix mixing
angles.

Thus, in a large class of mass matrices we have a natural suppression of
$d=5$ operator induced nucleon decay.

\vspace{1 cm}
{\bf Acknowledgments }

 \vspace{0.3cm}
We would like to thank G.Dvali for helpful conversations, which have
stimulated us to write this paper. We are also indebted to
J.L.Chkareuli for useful discussions. This work was supported in part
by International Science Foundation (ISF) under the Grant No.MXL000.


\begin{thebibliography}{99}
\bibitem{1}
Sakai N. and Yanagida T. // Nucl.Phys. 1982. V. B197. P. 533;
Weinberg S. // Phys. Rev. 1982. V. D26. P. 287;
Nath P., Chamseddine A.H. and  Arnowitt R. // Phys.Rev. 1985. V. D32. P.
2348.
\bibitem{2}
Campbell B., Elis J. and Nanopoulos D.V. // Phys.Lett. 1984. V. B141. P. 229;
Nath D. and Arnowitt R. // Phys.Rev. 1988. V. D38. P. 1479.
\bibitem{3}
Arnowitt R. and Nath P. // Phys.Rev.Lett. 1992. V. 69. P. 725;
Nath P. and Arnowitt R. // Phys.Lett. 1992. V. B287. P. 89.
\bibitem{4}
Dvali G.R. // Phys.Lett. 1992. V. B287. P. 101.
\bibitem{5}
Anderson G., Dimopoulos S.,
Hall L.J., Raby S. and Starkman G.D. // Phys.Rev. 1994. V. D49. P. 3660.
\bibitem{6}
Chadna S. and Daniel M. // Nucl.Phys. 1983. V. B229. P. 205.
\bibitem{7}
M.B.Gavela et.al. // Nucl.Phys. 1989. V. B132. P. 269.
\end{thebibliography}
\end{document}